\documentclass[a4paper]{jpconf}
\usepackage{graphicx}
\usepackage{graphics}
\usepackage{epstopdf}
\usepackage{subfigure}
\usepackage{amsmath,environ}
\usepackage{float}
\usepackage{color}
\begin{document}
\title{Time Dependent Study of Multiple Exciton Generation in Nanocrystal Quantum Dots}
\author{Fikeraddis A. Damtie}
\address{Mathematical Physics, Lund University}
\ead{Fikeraddis.Damtie@teorfys.lu.se}
\author{Andreas Wacker}
\address{Mathematical physics, Lund University.}
\ead{Andreas.Wacker@fysik.lu.se}
\begin{abstract}
We study the exciton dynamics in an optically excited nanocrystal quantum dot. 
Multiple exciton formation is more efficient in nanocrystal quantum dots compared to bulk semiconductors due to 
enhanced Coulomb interactions and the absence of conservation of momentum. 
The formation of multiple excitons is dependent on different excitation
parameters and the dissipation. We study this process within a
Lindblad quantum rate equation using the full many-particle states.
We optically excite the system by creating a single high energy 
exciton $E_{SX}$ in resonance to a double exciton $E_{DX}$. With Coulomb 
electron-electron interaction, the population can be transferred from the single exciton to the double 
exciton state by impact ionisation (inverse Auger process). The
  ratio between the recombination processes and the absorbed photons provide
  the yield of the structure. We observe a quantum yield of comparable value to experiment assuming typical 
experimental conditions for a $4$ nm PbS quantum dot.
\end{abstract}

\section{Introduction}
 Multiple exciton generation (MEG) is the process by which a single absorbed photon with high energy 
creates more than one electron hole pair ~\cite{BeardPCL2011,NozikCPL2008, ShabaevNL2006, SchallerPRL2004}. 
It allows for  converting the excess energy of absorbed photon into an
additional photocurrent. 
In conventional solar cells based on bulk semiconductors, the conversion efficiency is limited and most of 
the excess energy is lost as heat within picosecond time scale after absorption.   
Due to quantum confinement effects, such as, relaxed momentum conservation,
modified carrier-cooling rates, reduced dielectric screening of the quantum dot surface and enhanced 
Auger processes, an enhancement of MEG is expected in quantum dots compared to 
their bulk counterparts~\cite{BeardNL2010, KlimovARPC2007}.
Lead Chalcogenide QDs have been intensively studied in relation to MEG since the 
first demonstration of MEG in PbSe quantum dots ~\cite{SchallerPRL2004, WiseAccChemRes2000}.
These quantum dots are known to exhibit high confinement effects
combined with a small band gap of bulk materials, 
e.g. $0.41$ eV for PbS and 
$0.28$ eV for PbSe at room temperature~\cite{KangJOSA97}. 
For a small lead based quantum dot, a quantum yield of 
$1.21\pm0.05$ has been shown in a recent paper ~\cite{KarkiSciRep2013}
which means that the number of induced excitons surpassed the
  absorbed photons by about $20\%$.
Apart from lead based dots, Cadmium Chalcogenide quantum dots, indium based quantum dots (InAs and InP) and Silicon quantum
dots have been studied for exploring the efficiency of MEG in quantum dots ~\cite{SchallerAPL2005, GachetNL2010, 
PijpersJPC2007, StubbsPRB2010, BeardNL2007}.  
From an experimental perspective, very short time resolution and very low excitation fluence 
are two requirements that needs to be fulfilled when measuring MEG quantum yield~\cite{SmithNanomaterials2013}.  
The first requirement is due to the rapid Auger recombination ($\approx 10-100$ ps)~\cite{KlimovScience2011} for multi-excitons in 
quantum dots in comparison to lifetime of single excitons ($\approx 10-100$ ns)~\cite{CrookerAPL2003}. To be able to 
capture these fast processes, techniques with time resolution much faster 
than the recombination times is needed.  Ultrafast lasers can produce pulses as short 
as $100$ fs or less in duration which offer adequate time resolution needed \cite{tragerBook2007}. 
The second requirement is to avoid multi-exciton formation due to successive excitation of 
quantum dots with two or more photons within the life time of a single exciton.
The observation of multi-exciton recombination at low pump fluence can then be considered as the signature 
for MEG. 
Ultrafast transient absorption spectroscopy, which is a pump probe technique, is one of the mostly used 
experimental techniques for measuring MEG
  \cite{ChoiSciRep2013,KarkiSciRep2013,SemoninScience2011}. 
MEG in quantum dots has been simulated by several groups, with a 
strong focus on the impact ionisation process.  
Shabaev \textit{et al.} \cite{ShabaevNL2006} used a time dependent density matrix approach 
to study efficient MEG in nanocrystals as a coherent superposition of 
single and multi-exciton states and showed that efficient MEG is a 
consequence of suppressed thermalization rate of the initially excited single exciton.
Allan \textit{et al.} \cite{AllanPRB2006} showed that experimentally observed 
MEG yield in PbSe nanocrystals can be explained by impact ionisation process which is very efficient in PbSe and PbS 
nanocrystals with very fast ($\sim$ fs) lifetime for the single exciton to relax into the double exciton.
Beard \textit{et al.} \cite{BeardNL2010} compared MEG in nanocrystal with impact ionisation in bulk and showed
that the calculated quantum yield for nanocrystals is much larger than its
bulk counterpart. 
The effect of relaxation has been studied by Schulze \textit{et
  al.}~\cite{SchulzePRB2011}. In the same spirit, 
Azizi and Machnikowski~\cite{AziziPRB2013} showed that relaxation 
can even increase the yield under certain 
conditions in a phenomenological 
three-state model.

\begin{figure}
\centering
\hfill
\subfigure[Schematic single particle energies for the quantum dot considered showing 
the single exciton and double exciton configurations]{\includegraphics[width=0.4\textwidth]{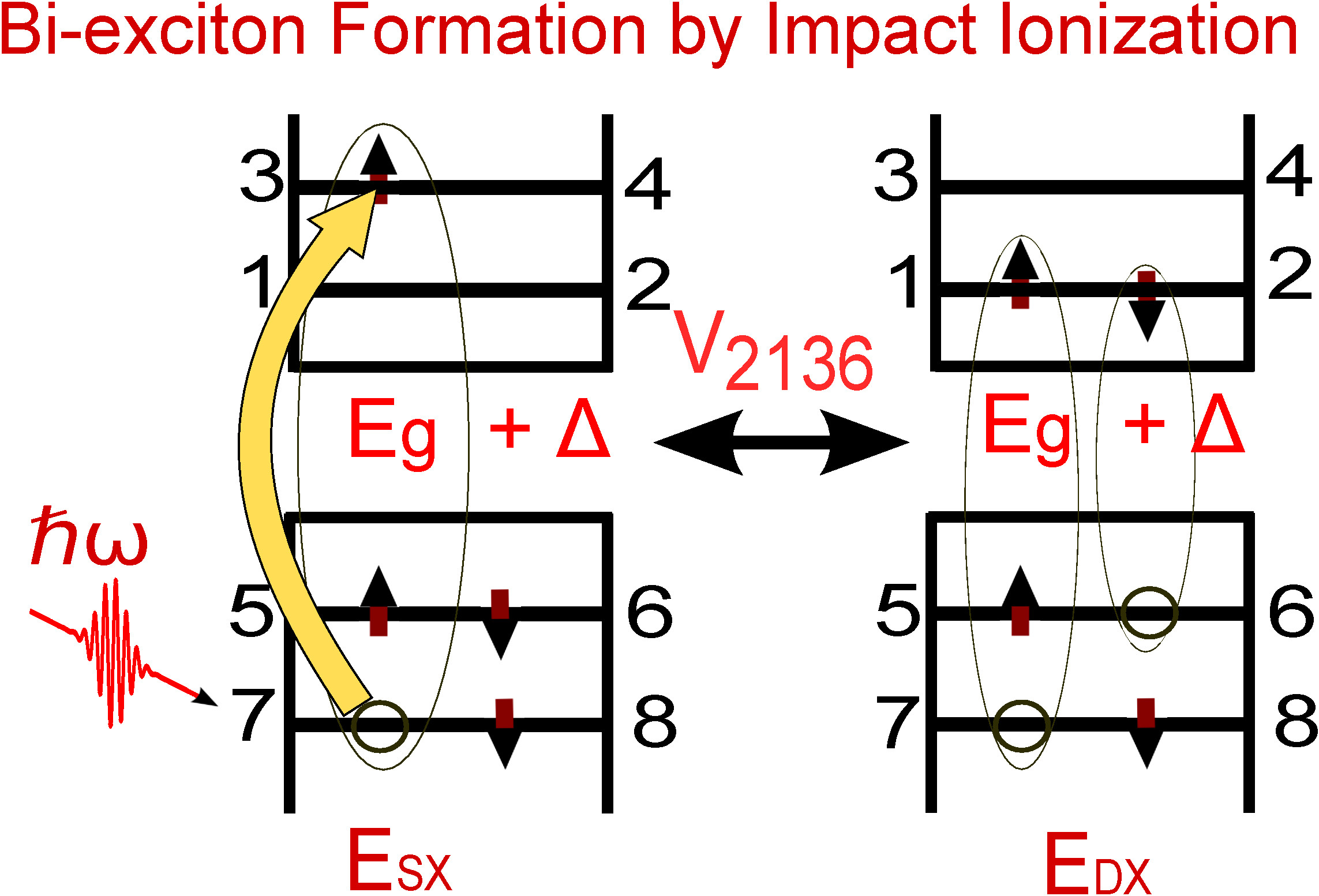}}
    \label{fig:subfig1}
\hfill
    \subfigure[Calculated energies of the high energy exciton $E_{SX}$ and 
double exciton $E_{DX}$ as a function of the parameter $\Delta$ taking into account all interactions.]{\includegraphics[width=0.49\textwidth]{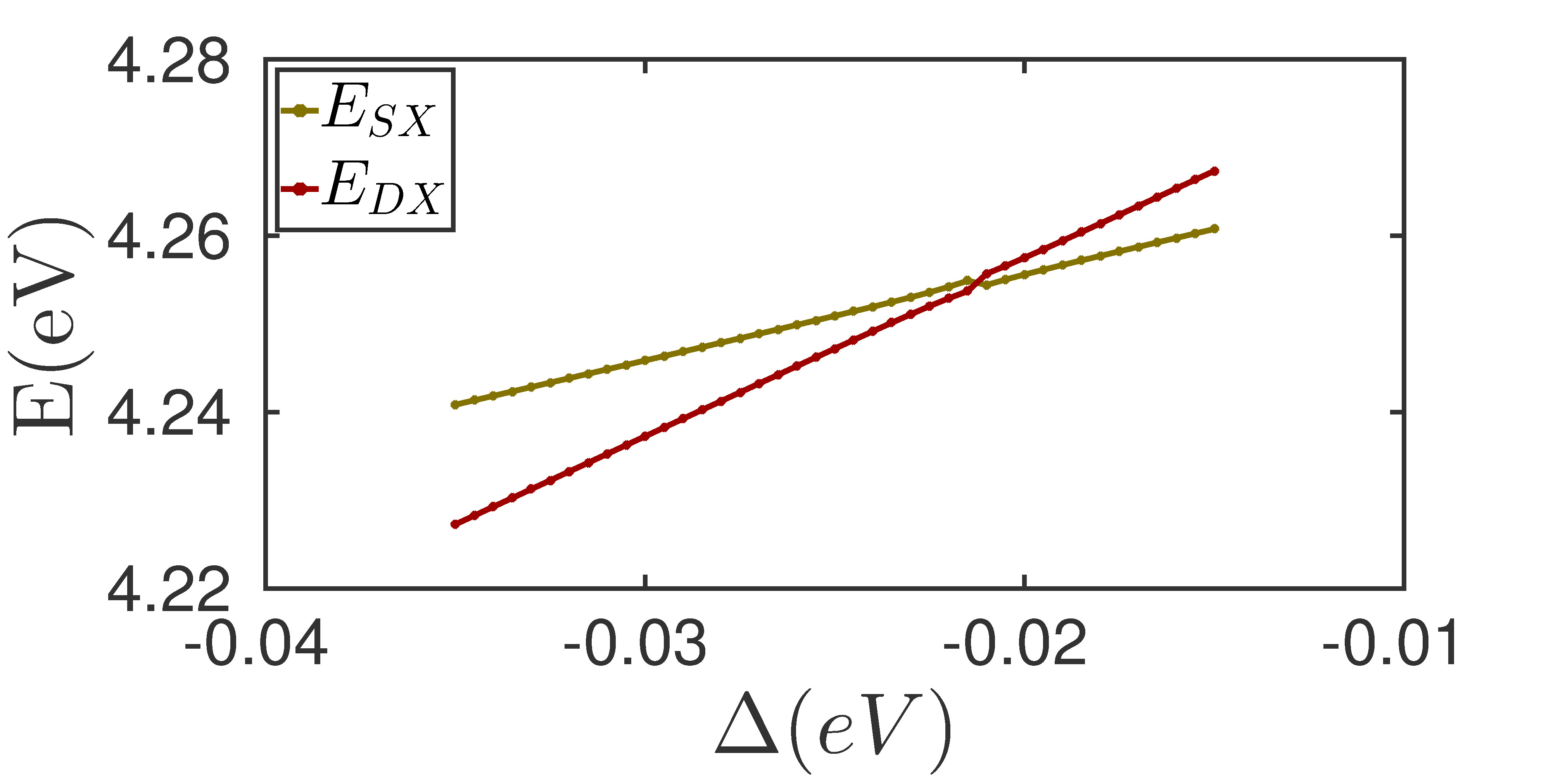}}
    \label{fig:subfig2}
\caption[Optional caption for list of figures]{Model studied and many particle energies of 
two important exciton states for the study}
\label{fig:Model and Energies}
\end{figure}

In this work, we  perform a quantum kinetic description based on 
a selected set of microscopic single particle states similar to 
\cite{SchulzePRB2011}. However, we work in the corresponding many-body 
basis, which allows for a better identification of the exciton states.  
We use specific parameters for the recent experiment
\cite{KarkiSciRep2013} with $4$ nm PbS dots, where  MEG is studied after
the excitation by a resonant pulse. This situation is schematically shown in Fig.~\ref{fig:Model and Energies}(a).
Within this method, the transient changes to the single exciton and double exciton states by the light pulse can be examined to see 
the various intermediate processes before the system relaxes to its ground state. 
Since the transient properties depend highly on the different parameters which can be controlled by suitable experimental setup, 
understanding these processes at the microscope level can greatly benefit for optimizing MEG experiments.
The method can be used for similar quantum dot systems for interpreting experimental measurements on MEG yield and transient 
absorption experiments.

\section{Method and Model}
In order to describe the quantum kinetics in a PbS quantum dot under irradiation by a short laser pulse, we construct a set of many-body basis states 
which are diagonal with respect to the electron-electron interaction. Within this basis we formulate a Lindblad kinetics to include relaxation, recombination and dephasing.
The different steps in the setup are addressed below.

\subsection{Single particle levels}\label{SecSingleParticle}
The energies of the single particle states for the spherical PbS quantum dot are evaluated using a 
K$\cdot$P method neglecting the multiplicity of valleys~\cite{KangJOSA97}.
As an input for the K$\cdot$P calculations, the material parameters for PbS quantum dot were 
adopted from Table 1 of the paper ~\cite{KangJOSA97}.  $m^+=m_e/3$ and $m^-=m_e/2.5$  are the conduction and 
valence band-edge effective masses respectively. The momentum-matrix element taken between the band-edge 
states of the conduction and valence band states is given by $P=\sqrt{2.5(eV)m_e/2}$. 
Here $m_e$ is the free electron mass. Furthermore, an energy gap $E_g=0.41$ eV at room temperature was also 
taken from the table. 
This gap is very narrow compared to most semiconductor materials which makes  
PbS quantum dots an excellent platform for studying MEG. 
\begin{table}[t]
\begin{center}
\begin{tabular}{|c|c|c|c|c|}
\hline
Orbital&n&l&$E^c_{nl}(eV)$&$E^v_{nl}(eV)$\\
\hline
1s&1&0 &$E_{1/2}=0.650280$&  $E_{5/6}=-0.6972889$\\
\hline
1p&1&1 &0.964789&-1.0609576\\
\hline
1d&1&2 & 1.319638& -1.4778528\\
\hline
2s&2&0 &1.482883& -1.6709188\\
\hline
1f&1&3 &1.721845&-1.954428\\
\hline
2p&2&1 &$E_{3/4}=1.99421$& $E_{7/8}=-2.27846$\\
\hline
1g&1&4 &2.174896& -2.493800\\
\hline
2d&2&2 &2.562865& -2.956857\\
\hline
\end{tabular} 
\caption{The lowest single-particle levels for the quantum dot considered in the study. The 1s and 2p levels are identified as most relevant and used in the 
calculations with the notation given in Fig.~\ref{fig:Model and Energies}(a)}
\label{Table:1}
\end{center}
\end{table}
In table~\ref{Table:1} the energies of the lowest single-particle levels of a $4$ nm PbS quantum dot 
are listed. 
The most relevant levels are the ground states (1s) of the conduction and valence band, which are the levels 1,2 and 5,6 in Fig.~\ref{fig:Model and Energies}(a), 
respectively. Here the different spin orientations ($\uparrow$ for odd numbers  and $\downarrow$ for even numbers) have been taken into account. In a single particle picture, 
MEG requires the energy difference between the excited and the ground state matching the gap $E_1-E_5$. The 2p conduction band level is the first to be close to this resonance. 
Thus this level, as well as it's corresponding valence band state (where optical transitions are in particular strong) are selected for the excited states 3,4 and 7,8. 
The precise shape of the quantum dots is not known and the K$\cdot$P modeling is not perfect for energies far beyond the gap. In order to mimic the the impact of such modifications, 
we consider a phenomenologically modified band gap $E_g+\Delta$ in the following. The detuning parameter $\Delta$ shifts all 
the conduction band levels $E_{nl}^c$ as listed in Table \ref{Table:1}. By varying the $\Delta$, the resonance between the $|SX\rangle$ 
and $|DX\rangle$ states can be controlled which determine the probability of the population transfer between these two 
exciton states.
In experiment it can be achieved by changing the geometry of the quantum dots. Fig. 1b shows the many particle energies for different values of the detuning parameter.
As we go along the horizontal axis, the detuning parameter is increasing which has the effect of increasing the band gap. 
As a result, the many particle energies rearrange themselves to this new band gap energy. In the specific spectrum considered, for $\Delta=0$
the $|SX\rangle$ is higher in energy by $\approx 0.0215eV$ than the $|DX\rangle$ state. Thus, the resonance can be achieved at $\Delta=-0.0215eV$, 
which is an avoided crossing as shown in Fig. 1b.
 
\subsection{Coulomb matrix elements}
MEG by impact ionisation is resulting from the Coulomb electron-electron interaction which is highly enhanced in 
quantum dots due to the high overlap of states. Here we calculate  the corresponding matrix elements for a  cuboid geometry which has a  volume equivalent to the volume 
of a 4 nm spherical quantum dot. This simplifies the calculations with wave functions $\sin(n_x\pi x/L_x)\sin(n_y\pi y/L_y)\sin(n_z\pi z/L_z)$. Here we identify the 1s state with 
$(n_x,n_y,n_z)=(1,1,1)$ and the 2p state with $(1,1,4)$, on the basis of their parity and the level sequence for the respective geometries. We estimate that the corresponding error 
is comparable to others resulting from the simplified screening with constant $\epsilon$ and the use of envelope functions.
The general expression for the Coulomb matrix element is 
given by the integral:
\begin{equation}\label{Eq:matrixelement}
 V_{n_{1}n_{2}m_{1}m_{2}}=\frac{e^{2}}{4\pi \epsilon\epsilon_{0}}\int\int d^{3}\textbf{r}d^{3}\textbf{r'} \phi^{*}_{n_{1}}
 (\textbf{r})\phi^{*}_{n_{2}}(\textbf{r'})\frac{e^{-\lambda |\textbf{r}-\textbf{r'}|}}{|\textbf{r}-\textbf{r'}|}\phi_{m_{1}}
 (\textbf{r'})\phi_{m_{2}}(\textbf{r})
\end{equation}
which enter the Hamiltonian in occupation number representation as
\begin{equation}
 \hat{H}_{ee}=\sum_{n_2<n_1,m_1<m_2}V^{\mathrm{eff}}_{n_{1}n_{2}m_{1}m_{2}}\hat{a}_{n_1}^\dag\hat{a}_{n_2}^\dag\hat{a}_{m_1}\hat{a}_{m_2}\quad\textrm{with }V^{\mathrm{eff}}_{n_{1}n_{2}m_{1}m_{2}}=
 V_{n_{1}n_{2}m_{1}m_{2}}-V_{n_{1}n_{2}m_{2}m_{1}}
\end{equation}
In Eq.(\ref{Eq:matrixelement}) a screening length $\lambda^{-1} \approx 5L_z$ was assumed which is 
 a length much larger than the dimension of the dot. This has no significant effect on  the overall result of the integral.
 As it will be described below, the evaluation of the Coulomb matrix element involve an integral in 
 Fourier space where $\lambda$ stabilizes the numerics.
 In addition, a relative permittivity of $\epsilon=17.2$ 
 was used~\cite{MadelungBookChap1998}.
The integral in Eq.(\ref{Eq:matrixelement}) is solved by splitting it into parts which involve $\textbf{r}$ 
and $\textbf{r'}$ and taking the Fourier transform of the screened Coulomb potential. 
This gives pairs of integrals in Fourier space of the form 
\begin{equation}\begin{split}\label{Eq:A_integrals}
A(\textbf{q})_{n_1,m_2}=&\int d^3\textbf{r}\phi_{n_1}^*(\textbf{r})\e^{i\textbf{q}\cdot\textbf{r}}\phi_{m_2}(\textbf{r}) \\
A^*(\textbf{q})_{m_1,n_2}=&\int d^3\textbf{r'}\phi_{n_2}^*(\textbf{r'})\e^{-i\textbf{q}\cdot\textbf{r'}}\phi_{m_1}(\textbf{r'}) \\
\end{split}\end{equation}
which contain pairs of indices $(n_1,m_2)$ and $(n_2,m_1)$ and $\textbf{q}$ is the wave vector.
Generally the Coulomb matrix elements can involve envelope functions $\phi_i$ either all from 
the same band or at least one is from the different band. If the pairs, $(n_1,m_2)$ originate from the same band, 
the the integrals in Eq.(\ref{Eq:A_integrals}) can be evaluated directly by using the envelope functions. 
However if they originate from different bands, the evaluation is more complicated. In this case,
one has to consider a multi-band model. In this work, a two band model \cite{SirtoriPRB1994} 
was used for evaluating the integral which involve the conduction and valence bands.  
The result gives a prefactor times the integral of the corresponding intraband pair 
with this prefactor being a constant proportional to the interband dipole matrix element times the wave vector $\mathbf{q}$.
Two types of electron-electron interactions dominate.  
Direct terms of the type $V^{\mathrm{eff}}_{n_{1}n_{2}n_{2}n_{1}}$ which account for direct electron-electron
 interaction within the same band (intraband) e.g $V^{\mathrm{eff}}_{2112}$ or with different bands (interband) e.g. $V^{\mathrm{eff}}_{5115}$.
 These direct terms are the strongest ones with calculated values of the order of $\approx 100$ meV.  
 The other types of elements which are of interest are the scattering terms $V^{\mathrm{eff}}_{n_{1}n_{2}m_{1}m_{2}}$ 
which account for several different types of interactions. The matrix element $V^{\mathrm{eff}}_{1263}$,
 which couples the single exciton
and the double exciton for MEG (Auger term), is of the order of $\approx -0.6$ meV. 

\subsection{Interaction with light and dipole matrix elements}
The coupling to the laser field is described by the Hamiltonian 
\begin{equation}
 \hat{H}_{I}(t)=\sum_{n,m}e E(t) \langle n |  \hat{z} | m\rangle \hat{a}_{n}^\dag\hat{a}_{m}
 \end{equation}
where $e$ is the elementary charge, and $E(t)$ the electrical component of the laser field, which is assumed to be z-polarised.

The dipole matrix elements $\langle n |  \hat{z} | m\rangle$ are divided into two groups: 
interband and intraband matrix elements.  
The interband dipole matrix elements are essentially determined by the z-matrix element of the lattice periodic functions and
 diagonal in the spatial structure of the envelope functions \cite{DaviesBook1997}. On the basis of 
 the p-matrix element between the valence and conduction band states 
 from \cite{KangJOSA97} we find
\begin{equation}\label{Eq:Dipole_Elements_6}
\langle 1|\hat{z}|5\rangle =\langle 3|\hat{z}|7\rangle =\langle 2|\hat{z}|6\rangle =\langle 4|\hat{z}|8\rangle= 0.7371 \;\mathrm{nm}
\end{equation} 
For the intraband case, the matrix element is evaluated by the envelope functions. 
Using the single particle wavefunctions of cuboid quantum dots we find
\begin{equation}\begin{split}\label{Eq:Dipole_Elements_10}
 \langle 3|\hat{z}|1\rangle =\langle 7|\hat{z}|5\rangle= \langle 4|\hat{z}|2\rangle =\langle 8|\hat{z}|6\rangle 
 =-0.05764\;\mathrm{nm}
\end{split}\end{equation}
A similar result ($0.045$ nm) is obtained from the 1s and 2p eigenstates of the spherical dot in a single band approximation, which 
shows that the cuboid states are a reasonable approximation.
For PbSe quantum dots, An \textit{et al.} \cite{AnNL2006} calculated the electronic band structure using 
an atomistic pseudopotential method which indicated that care has to be taken when using K$\cdot$P method as it 
misses some optical selection rules.

The oscillating electric field is assumed to have  a Gaussian envelope 
\begin{equation}
 E(t)=E_0 \exp\left(-\frac{(t-t_0)^2}{\tau^2}\right) \sin(\omega(t-t_0))
\end{equation}
where in the above expression 
$\omega$ is the central frequency for the pulse which can be tuned to the resonance frequencies.
$\tau$ describes the width of the pulse and $E_0$ is the electric field amplitude. 
Here we parametrise $E_0$ by the pulse area
\begin{equation}
 \Theta=\frac{e}{\hbar}E_0\langle 1|\hat{z}|5\rangle \int_{-\infty}^{\infty} dt\, \exp\left(-\frac{(t-t_0)^2}{\tau^2}\right)
\end{equation}
In a two two-level system, a $\Theta=\pi$ pulse flips the ground state to the excited~\cite{MukamelBook1995}.
From the experimental pulse parameters given in the supplementary information of~\cite{KarkiSciRep2013}
a pulse strength of the order of $\Theta\approx 1$ is possible. However for the calculations in this work, a value of 
$\Theta$ smaller than one is used to be able to achieve the low excitation 
fluence requirement~\cite{SmithNanomaterials2013} which is about $1\mu J$ per pulse for the sample assumed ~\cite{KarkiSciRep2013}.  

\subsection{Many particle states}
The total Hamiltonian for the system can be divided into two main parts, time independent $\hat{H}_0$ 
and time dependent interaction part $\hat{H}_I(t)$. 
\begin{equation}
\hat{H}_{\mathrm{eff}}(t)=\underbrace{\sum_i E_i \hat{a}_i^\dag\hat{a}_i +\hat{H}_{ee}}_{\hat{H}_0}+\hat{H}_I(t)
\label{Eq:Total Hamiltonian}
\end{equation}
The many body states are obtained by exact diagonalisation of  $\hat{H}_0$. Fixing the total number of electrons to 4, the 4 particle 
many-body ground state is where all the levels in the conduction band are 
empty and all the levels in the valence band are occupied. 
In Fock state representation it is given by 
\begin{equation}
|GS\rangle\approx |\underbrace{0 0 0 0}_{c} \underbrace{1 1 1 1}_{v} \rangle 
\end{equation}
where the occupation numbers $|n_1,n_2,n_3,n_4,n_5,n_6,n_7,n_8\rangle $ relate to the 8 states in 
Fig.~\ref{fig:Model and Energies}(a).
The lowest energy excitation $|ex\rangle$ is due to an electronic transition across 
the band gap from the highest level in the valence band to the lowest level in the conduction band, which is called exciton. 
For the system considered, the energy of the lowest singlet excitation is $E_{ex}=1.334$ eV above the 
ground state.
The corresponding many body eigenstate is $|ex\rangle\approx
1/\sqrt{2}(|1 0 0 0 0 1 1 1\rangle - |0 1 0 0 1 0 1 1\rangle)$. 
The energy $E_{ex}$ is slightly smaller than the difference between the single-particle levels 1 and 5 (1.347 eV from
 table \ref{Table:1}) due to small admixtures of other states. (The experimentally measured gap 
 of 1.07 meV is even smaller. This might be due a different shape, but the admixture from lower lying levels (e.g. 1p) neglected in our model will also play a role.)
Two other many-body states play a very important role in the description of MEG:  The high energy 
single exciton $|SX\rangle\approx 1/\sqrt{2}(|0 0 1 0 1 1 0 1\rangle - |0 0 0 1 1 1 1 0\rangle)$
and the double exciton  $|DX\rangle\approx 1/\sqrt{2}(|1 1 0 0 1 0 0 1\rangle - |1 1 0 0 0 1 1 0\rangle)$ which are both 
singlets and close in energy. The detuning between these two states can be controlled by the parameter $\Delta$
(which modifies the band gap, see section 
\ref{SecSingleParticle}) as shown in Fig.~\ref{fig:Model and Energies}(b).  At $\Delta\approx-0.0215$ eV the two states show an anticrossing with 
the difference $|E_{DX}-E_{SX}|=2V_{DX,SX}$, where $|V_{DX,SX}|\approx0.6$ meV  is the Coulomb matrix element between the states $|DX\rangle$ and $|SX\rangle$.   

\subsection{Equation of motion for the density matrix}
The system consists of a classical light field and a quantum system with discrete energies. The change 
to the system induced by the light field can be best described by the time evolution of the 
reduced density operator for the system given by the Lindblad equation~\cite{LindbladCMP1976}. 
\begin{equation}
\hbar\frac{d}{dt}\hat{\rho}_S(t)=i[\hat{\rho}_S(t),\hat{H}_{\mathrm{eff}}(t)]+\sum_{j=1}^{N_{\mathrm{jump}}}\Gamma_j
\bigg[\hat{L}_j\hat{\rho}_S(t)\hat{L}^{\dag}_j-\frac{1}{2}(\hat{\rho}_S(t)\hat{L}^{\dag}_j\hat{L}_j
+\hat{L}^{\dag}_j\hat{L}_j\hat{\rho}_S(t))\bigg]\label{EqLindblad}
\end{equation}
The time evolution in Eq.~(\ref{EqLindblad}) was solved in the many particle basis with a 4th order Runge-Kutta method.
The jump operators $\hat{L}_j$ describe
different dissipation processes (with rate $\Gamma_j
/\hbar$) which are restoring the ground state at sufficiently long 
time after the excitation. Here we use:\\
Relaxation in the conduction band 
 \begin{equation}
 \hat{L}_{\mathrm{rel}}=\hat{a}^{\dag}_{1\uparrow}\hat{a}_{3\uparrow} + \hat{a}^{\dag}_{2\downarrow}\hat{a}_{4\downarrow} 
 \;\;\;\;\; \mbox{with strength}  \;\;\;\;\; \Gamma_\mathrm{Relaxation}
 \end{equation}
Relaxation in the valence band
 \begin{equation}
 \hat{L}_{\mathrm{rel}}=\hat{a}^{\dag}_{7\uparrow}\hat{a}_{5\uparrow} + \hat{a}^{\dag}_{4\downarrow}\hat{a}_{8\downarrow} 
 \;\;\;\;\; \mbox{with strength}   \;\;\;\;\; \Gamma_\mathrm{Relaxation}
 \end{equation}
Recombination across the band gap
 \begin{equation}
 \hat{L}_{\mathrm{rec}}=\hat{a}^{\dag}_{5\uparrow}\hat{a}_{1\uparrow} + \hat{a}^{\dag}_{6\downarrow}\hat{a}_{2\downarrow} 
 \;\;\;\;\; \mbox{with strength} \;\;\;\;\; \Gamma_\mathrm{Recombination}
 \end{equation}
Dephasing in the conduction band
 \begin{equation}\begin{split}
& \hat{L}_{\mathrm{deph}}=\hat{a}^{\dag}_{1\uparrow}\hat{a}_{1\uparrow} + \hat{a}^{\dag}_{2\downarrow}\hat{a}_{2\downarrow} 
 \;\;\;\;\; \mbox{with strength} \;\;\;\;\; \Gamma_\mathrm{Dephasing}\\
&  \hat{L}_{\mathrm{deph}}=\hat{a}^{\dag}_{3\uparrow}\hat{a}_{3\uparrow} + \hat{a}^{\dag}_{4\downarrow}\hat{a}_{4\downarrow} 
 \;\;\;\;\; \mbox{with strength} \;\;\;\;\; \Gamma_\mathrm{Dephasing}\\
\end{split}\end{equation}
Dephasing in the valence band 
 \begin{equation}\begin{split}
& \hat{L}_{\mathrm{deph}}=\hat{a}^{\dag}_{5\uparrow}\hat{a}_{5\uparrow} + \hat{a}^{\dag}_{6\downarrow}\hat{a}_{6\downarrow} 
 \;\;\;\;\; \mbox{with strength} \;\;\;\;\; \Gamma_\mathrm{Dephasing}\\
&  \hat{L}_{\mathrm{deph}}=\hat{a}^{\dag}_{7\uparrow}\hat{a}_{7\uparrow} + \hat{a}^{\dag}_{8\downarrow}\hat{a}_{8\downarrow} 
 \;\;\;\;\; \mbox{with strength} \;\;\;\;\; \Gamma_\mathrm{Dephasing}\\
\end{split}\end{equation}
These jump operators are defined in such a way that
they all conserve the total spin indicated by the spin indices. 
The different decoherence mechanism which were assumed phenomenologically in Eq.~(\ref{EqLindblad}) can be associated to all forms of intrinsic 
scattering mechanisms other than electron-electron scattering, which has already been included in the effective Hamiltonian.

\section{Results}
We simulate Eq.~(\ref{EqLindblad})  in order to determine the density matrix as a function of time. 
In all calculations we use $\tau=150$ fs and $t_0=200$ fs together with 
the initial condition $\hat{\rho}=|GS\rangle\langle GS|$ at $t=0$. Regarding the dissipation mechanisms,
dephasing is the fastest scale and we use
$\Gamma_{\mathrm{Dephasing}}=6$ meV, which corresponds to a $\tau_{\mathrm{Dephasing}}=\hbar/\Gamma_{\mathrm{Dephasing}}\approx100$ fs.
Relaxation is typically slightly slower, and we use $\Gamma_{\mathrm{Relaxation}}=3.3$ meV. 
These are typical values in solids. Recombination is typically on the ns time scale. In order to save simulation time,
we use a larger value of  $\Gamma_{\mathrm{Recombination}}=0.33$ meV (corresponding to 
$\tau_{\mathrm{Recombination}}=\hbar/\Gamma_{\mathrm{Recombination}}\approx2$ ps), which is still a factor of $10$ slower than the other processes.

\begin{figure}
\hfill
\subfigure[Calculated number of recombinations]{\includegraphics[width=0.49\textwidth]{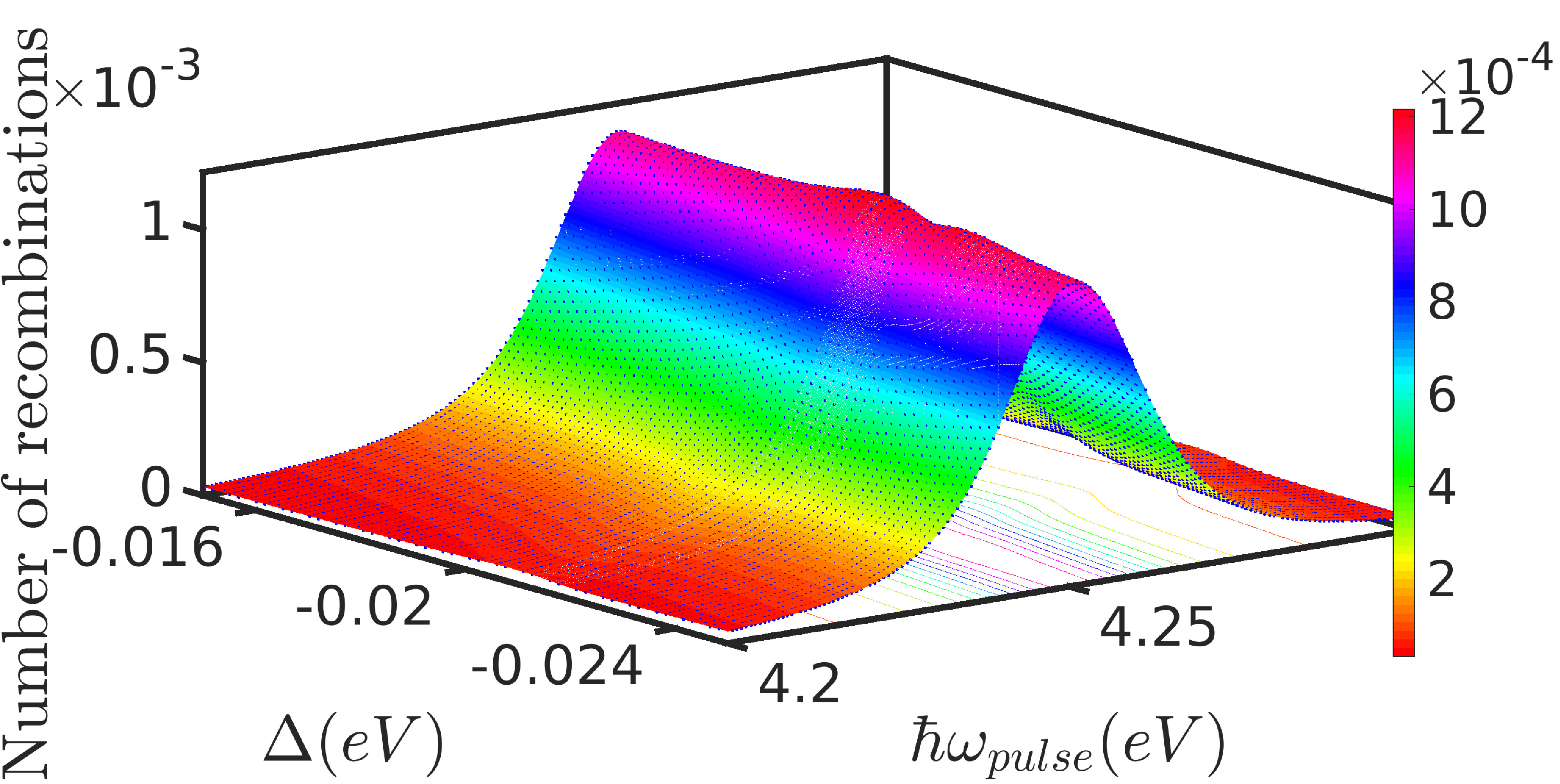}}
    \label{fig:Rate1}
\hfill
    \subfigure[Calculated absorbed energy]{\includegraphics[width=0.49\textwidth]{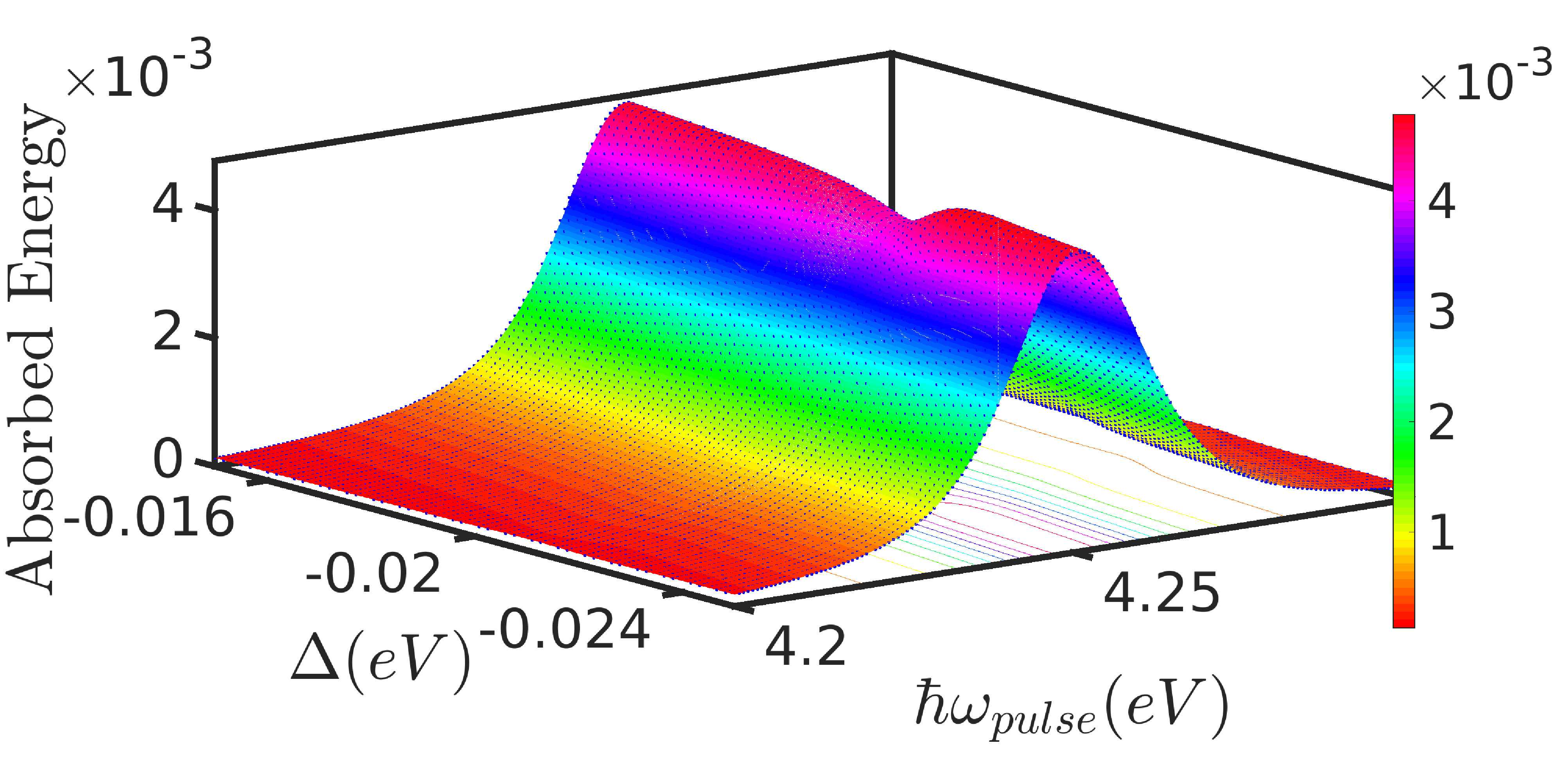}}
    \label{fig:Power1}
\caption[]{Calculated number of recombinations (a) and absorbed energy (b) as a function of $\Delta$ and pulse 
frequency close to the transition between the 2p states. 
The pulse area is $\Theta=0.035\pi$.}
\label{fig:subfigure_Rate_Power}
\end{figure}

Our main observable is the recombination rate
\begin{equation}
 \mathrm{Rec(t)}=\Gamma_{\mathrm{Recombination}}Tr\{\hat{L}_{\mathrm{rec}}\hat{\rho}\hat{L}_{\mathrm{rec}}^{\dag}\}
\end{equation}
Integrating over time provides the number of recombined electron-hole pairs generated by the pulse.
Fig.~\ref{fig:subfigure_Rate_Power}(a) shows the calculated number of recombinations
for excitations in the vicinity of the $2p$ resonance for different
values of $\Delta$. The average full width at half maximum (FWHM) of the peaks is about $\approx 22$ meV which is mainly due to the pulse width $\tau=150$ fs. 
It can be seen that recombination is enhanced when the two excitons 
are close in energy near $\Delta=-0.0215$ eV. The other ingredient needed for calculating the yield is the power 
transferred to the system by the oscillating field. It can be defined as the absorbed energy per unit time and 
can be calculated as  
\begin{equation}
 \mathrm{P(t)}=\frac{d}{dt}\langle H_0 \rangle=\frac{i}{\hbar}\langle [H_{\mathrm{eff}}(t),H_0] \rangle=
 \frac{i}{\hbar}\langle [H_I(t),H_0] \rangle
\end{equation}
Integrating over time provides the absorbed energy, which is displayed in
Fig.~\ref{fig:subfigure_Rate_Power}(b). This has a similar shape, except for a small dip near $\Delta=-0.0215$ eV.

\begin{figure}[H]
 \includegraphics[scale=.28]{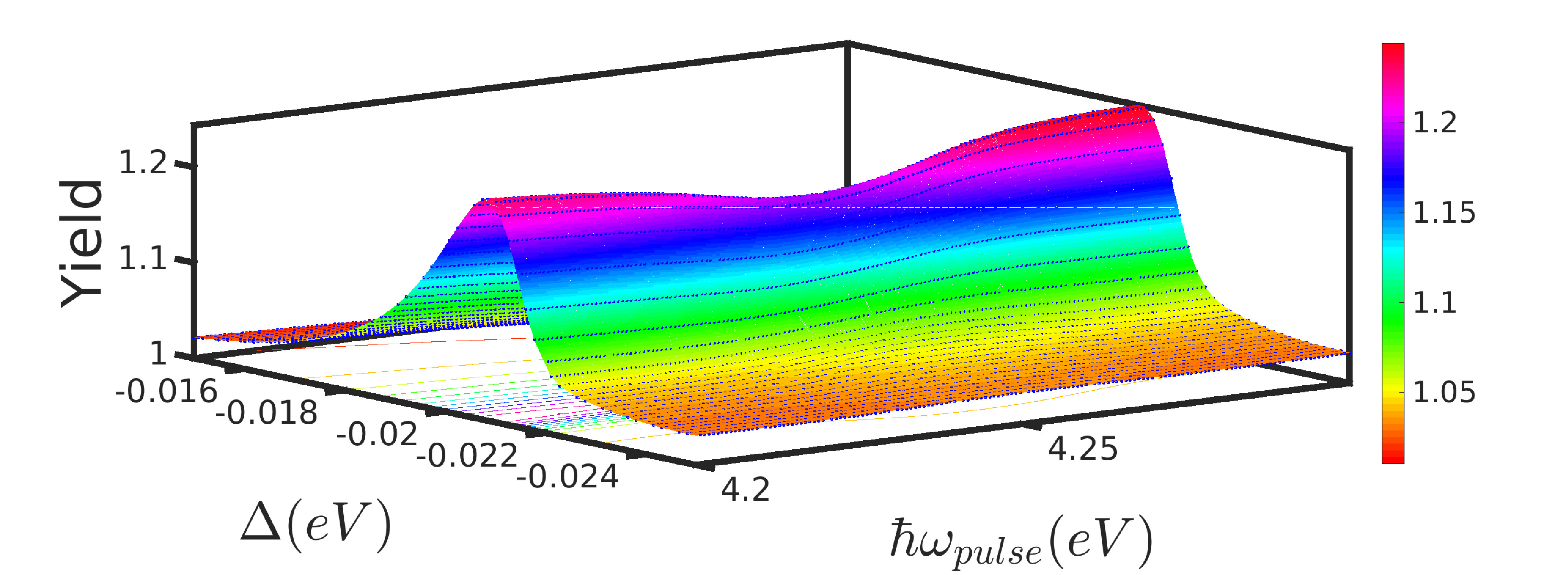}
 \caption{Quantum yield combining the data of Figs.~\ref{fig:subfigure_Rate_Power}(a,b)}
\label{Fig:Yield_1}
\end{figure}

The yield is then defined as the ratio between the number of recombinations and the absorbed photons.
The latter is given by the absorbed energy  divided by the photon energy. Due to the short pulse width the
photon energy is not well defined. Thus we take the energy $\hbar\omega_{\rm pulse}$, where the 
number of recombinations peaks for a given detuning $\Delta$. The corresponding result is plotted in 
Fig.~\ref{Fig:Yield_1}

We observe a quantum yield of $\approx 1.2$ close to resonance $\Delta\approx -0.0215$ eV. This yield agrees
well with the experimentally observed values in these dots reported by Karki \textit{et al.}\cite{KarkiSciRep2013}. 
However we obtain this number only for a specific type of dots, characterised by our phenomenological parameter 
$\Delta$.

Now we want to study the kinetics in detail
for $\Delta=-0.0215$ eV, where the single exciton and double exciton states are very close in energy.
Thus, the states $|SX\rangle$ and $|DX\rangle$ mix significantly, which results in the singlet states 
$|22\rangle$ and $|26\rangle$ in our many-body calculation. Analysing these states they can be decomposed
as
\begin{equation}\label{Eq:psi_22}
 |22\rangle\approx\alpha_1|SX\rangle + \beta_1|DX\rangle
\end{equation}
and 
\begin{equation}\label{Eq:psi_26}
|26\rangle\approx\alpha_2|SX\rangle + \beta_2|DX\rangle  
\end{equation}
where further contributions are neglected.
In terms of these coefficients the probability for the double exciton and single exciton is given by 

\begin{equation}\begin{split}\label{Eq:rho_DX}
&\rho_{\mathrm{DX,DX}}=\langle DX|\rho |DX\rangle=\alpha^2_1 \rho_{26,26}  
+  \alpha^2_2\rho_{22,22}- 2\alpha_1\alpha_2 \Re\{\rho_{22,26}\}\\
\end{split}\end{equation}

\begin{equation}\begin{split}\label{Eq:rho_SX}
&\rho_{SX,SX}=\langle SE|\rho |SX\rangle=\beta^2_1 \rho_{26,26}  
+  \beta^2_2\rho_{22,22}- 2\beta_1\beta_2 \Re \{\rho_{22,26}\}\\
\end{split}\end{equation}
respectively where $\Re$ denote the real part of the coherences between the states $|22\rangle$ and $|26\rangle$.

\begin{figure}[t]
 \includegraphics[scale=.28]{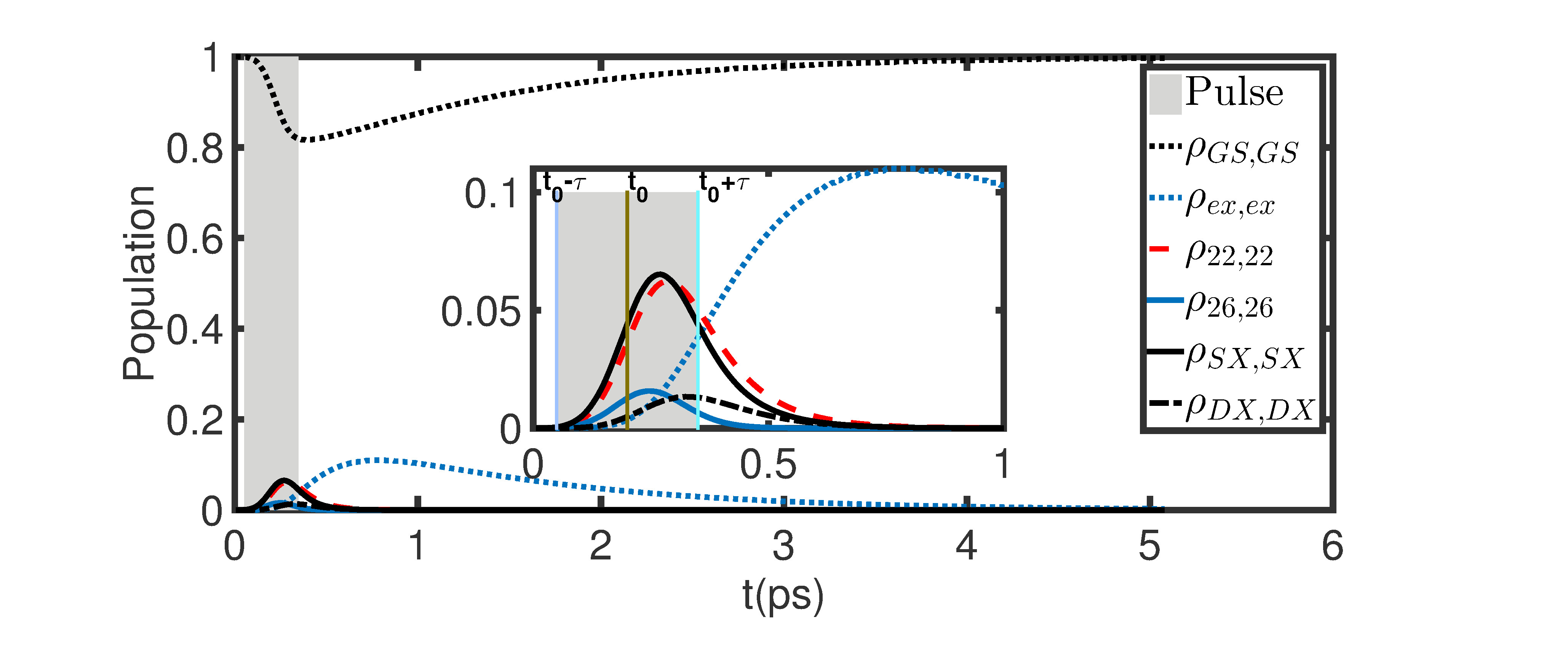}
 \caption{Time evolution for selected many particle states for $\Delta=-0.0215$ eV , $\hbar\omega_\mathrm{pulse}=4.255$ eV. An increased pulse area of $\Theta=\pi/2$ 
 is used in order to emphasise the effect. The grey area depicts the duration of the pulse. The inset is an enlargement for small times.}
\label{Fig:Dynamics_1}
\end{figure}

Fig.~\ref{Fig:Dynamics_1} shows the time evolution of some selected many-particle states when the 
two excitons are close and a pulse in resonance with the state $|SX\rangle$ is used. 
In the region where the pulse is 
active, coherent population transfer is dominating mechanism.
It can be seen that for a strong pulse $\Theta=\pi/2$ the system is excited significantly.
The $|SX\rangle$ state is the first to be excited as the pulse is tuned to be in resonance with 
this state and the transition is optically allowed. 
Subsequently (around $t_0$) the state $|DX\rangle$ starts to grow, which can be understood as impact ionisation due to the Coulomb interaction sketched in 
Fig.~\ref{fig:Model and Energies}(a). Almost at the same time, the single exciton ground states $|ex\rangle$ starts growing as it is fed by relaxation processes 
from the excited exciton $|SX\rangle$ via intermediate states not plotted here. Finally, the $|SX\rangle$ and  $|DX\rangle$ states lose their population after the 
pulse due to the slow recombination rate. In addition there is an Auger contribution, where $|DX\rangle$ is transferred back to $|SX\rangle$. 

Within the many-body eigenstates, the same story is told differently: The optically excited state $|SX\rangle$ is a linear combination of $|22\rangle$ and  $|26\rangle$. 
Thus the population of these two states is evolving in parallel. However, due to the energy splitting $\Delta \geq 2V_{DX,SX}$, the coherence $\rho_{22,26}$ is oscillating 
in time with the angular frequency $\Delta/\hbar$. Eqs.~(\ref{Eq:rho_DX},\ref{Eq:rho_SX}) show that this oscillation implies population changes in $|SX\rangle$ and  $|DX\rangle$. 
Relaxation processes reduce the populations of $|SX\rangle$ and  $|DX\rangle$ slightly after the peak of the pulse at $t_0=0.2$ ps was reached and the generation becomes weaker. 
Due to this relaxation and, equally important, dephasing, the oscillation period is never completed for the parameters used.

\section{Conclusion and outlook}
In this work, we have studied MEG due to optical excitations in $4$ nm PbS quantum dots. We took into account the electron-electron interaction by diagonalising the 
Hamiltonian. In this basis we solved the  equation of motion for the density matrix in the Lindblad form.
We find a yield of $1.2$ for a $\hbar\omega_{\rm pulse}\approx 4.2$ eV. This goes well with the experimental
observation of a yield of $1.2$ for photon energies surpassing 3 times the lowest exciton absorption \cite{KarkiSciRep2013}. 
Furthermore, we clarified the relation between the 
picture of impact-ionisation and a description with diagonal many-particle states.

While the values of the yield are promising, it is not clear, why these values for a particular range of parameters $\Delta$ are representative for the actual experiment. 
One might argue that the multitude of many-particle states arising if further single-particle levels are taken into account for, offers a plentitude of such resonances, 
so that they become generic similar to the scenario discussed for tunneling in \cite{GoldozianPreprint2015}. This has, however, to be studied further in detail. 
Another interesting point for further study is the optimisation of the yield and the harvesting of multiple excitons by appropriate contacts.
To be able to benefit from MEG, an efficient extraction mechanism is a key. One promising demonstration of 
efficiently collecting the multiple excitons produced is by Sambur \textit{et al.} \cite{Sambur01102010} 
in which they attached a single layer of PbS quantum dot to a thin oxide ($TiO_2$). 
They claim that the strong electronic coupling and favorable energy level alignment between the quantum dot 
and the oxide facilitates a quick extraction of multiple excitons before they recombine. 
This can actually be quantified using an extension of the model used
here. Corresponding work is ongoing.
\section{Acknowledgements}
We thank T\"onu Pullerits and Khadga Karki for helpful discussions and the Knut and Alice Wallenberg foundation as well as NanoLund for financial support.
\section{References}
\providecommand{\newblock}{}

\end{document}